\documentclass[12pt]{iopart}
\pdfoutput=1
\expandafter\let\csname equation*\endcsname\relax
\expandafter\let\csname endequation*\endcsname\relax
\usepackage[numbers,sort&compress]{natbib}
\usepackage{amsmath,amssymb,eucal,graphicx,subfigure}
\usepackage{eso-pic,calc}
\usepackage{graphicx}
\usepackage{epsfig}
\usepackage{bm}
\usepackage{color}
\usepackage{wasysym}

\usepackage[colorlinks=True]{hyperref}  
\hypersetup{
 	colorlinks=true,  
 	linkcolor=blue,  
 	citecolor=blue, 
 	filecolor=magenta,   
 	urlcolor=blue         
 }

\def\beqr{\begin{eqnarray}}
	\def\eqnr{\end{eqnarray}}
\def\beq{\begin{equation}}
	\def\bc{\begin{center}}
		\def\ec{\end{center}}
	\def\eqn{\end{equation}}

\begin{document}
\title{ $1/f^{\alpha}$ noise in the Robin Hood model}

\author{Abha Singh}
\address{Department of Physics, Institute of Science,  Banaras Hindu University, Varanasi 221 005, India}

\author{Rahul Chhimpa}
\address{Department of Physics, Institute of Science,  Banaras Hindu University, Varanasi 221 005, India}

\author{Avinash Chand Yadav\footnote{jnu.avinash@gmail.com}}
\address{Department of Physics, Institute of Science,  Banaras Hindu University, Varanasi 221 005, India}

\begin{abstract}
{We consider the Robin Hood dynamics, a one-dimensional extremal self-organized critical model that describes the evolution of low-temperature creep. One of the key quantities is the time evolution of the state variable (force noise). To understand the temporal correlations, we compute the power spectra of the local force fluctuations and apply finite-size scaling to get scaling functions and critical exponents. We find a signature of the $1/f^{\alpha}$ noise for the local force with a nontrivial value of the spectral exponent $0< \alpha < 2$. We also examine temporal fluctuations in the position of the extremal site and a local activity signal. We present results for different local interaction rules of the model. }
\end{abstract}

\maketitle
\section{Introduction}
Diverse temporal noises can exhibit low-frequency $1/f^{\alpha}$ noise in the power spectral density~\cite{RevModPhys.53.497}. Generally, the spectral exponent $\alpha$ lies between 0 (white) and 2 (Brownian noise). Examples vary, from voltage variations across a resistor to biological signals like DNA sequences~\cite{PhysRevLett.68.3805}.
Self-organized criticality (SOC), introduced by Bak-Tang-Wiesenfeld (BTW)~\cite{PhysRevLett.59.381, Christensen_2005, Pruessner_2012, MARKOVIC201441, Watkins2016} can explain the $1/f$ noise observed in non-equilibrium natural systems, although SOC is not a necessary condition. The SOC systems organize spontaneously into a critical state, where the long-range space-time correlation emerges naturally. A common feature of SOC is a power law in the avalanche size or duration distribution. The scaling property vanishes when the system's external driving rate is high. SOC seems to occur in diverse contexts, ranging from earthquakes~\cite{PhysRevLett.88.178501, PhysRevLett.114.088501} and biological evolution~\cite{PhysRevLett.73.906, PhysRevLett.71.4083, PhysRevLett.132.098402} to rainfall~\cite{ANDRADE1998557}. 

The BTW sandpile model exhibits the $1/f^{\alpha}$ noise in the avalanche activity signal monitored on a fast time scale~\cite{Laurson_2005, shapoval20241varphi}. Several variants of the BTW model also exhibit the $1/f$ noise. The total mass or energy fluctuations recorded at a slow (drive) time scale also show $1/f$ noise~\cite{PhysRevLett.83.2449, PhysRevLett.82.472, PhysRevE.85.061114,  PhysRevE.66.050101, PhysRevE.104.064132, Yadav_2022, Kumar_2022}. The Bak-Sneppen (BS) evolution model displays the $1/f$ noise in local activity~\cite{PhysRevLett.73.2162} and the number of species below a threshold~\cite{PhysRevE.53.4723, PhysRevE.63.063101}. 
Recently, the fitness fluctuations in the BS model have been shown to follow $1/f^{\alpha}$ noise with $\alpha \approx 1.2$~\cite{PhysRevE.108.044109}. The BS model demonstrates \emph{punctuated equilibrium}, wherein the long periods of stasis are interrupted by intermittent bursts. 

The naturally evolving system named low-temperature creep, or Robin Hood (RH) model, also exhibits the phenomenon of SOC~\cite{ZAITSEV1992411, article}. Commonly, creep refers to the evolution of a system under a constant external driving force. Originally proposed for dislocation movements, the RH model consists of a one-dimensional (1D) lattice, where a site $i$ at the time step $t$ has the height $h_i(t)$. At each time step, the site with the maximum height is selected: $h_m(t) = max \{h_i(t)\}$. The height evolves as $h_m(t+1) = h_m(t) - \Delta(t)$ and $h_{m\pm1}(t+1) = h_{m\pm1}(t) + \Delta(t)/2$, where $\Delta(t)$ is an independent random variable. For periodic boundary conditions, the total amount $\sum_{i=0}^{L} h_i(t)$ remains constant.

One can easily simulate the 1D RH model, starting with a flat interface. Selecting the first site to rob at random, all $L$ sites of the interface get updated at least once after $T$ steps. The average
$T$ for many independent runs scales with the system size $\langle T \rangle \sim L^{\mathcal{D}}$~\cite{PhysRevE.74.066110}. The avalanche size follows the power-law distribution $P(S) \sim S^{-\tau}$. To define an avalanche, consider the sites that are above a threshold height $h_0$ as live sites. An avalanche of threshold $h_0$ is the number of time steps $S$ in which the maximum height is greater than $h_0$. In extremal SOC models, various critical exponents can be typically expressed in terms of two important exponents: the avalanche dimension $\mathcal{D}$ and the correlation exponent $\nu$.
The exponents $\tau$ and $\nu$ satisfy a scaling relation
$\tau = 1+\left({\rm D}-1/\nu\right)/ \mathcal{D}$~\cite{PhysRevE.53.414}. The avalanche dimension $\mathcal{D}$ and the interface roughness exponent $\chi$ follow a scaling relation $\chi+{\rm D} = \mathcal{D}$.
In the 1D RH model, the critical exponents $\mathcal{D} = 2.23$ and $\tau = 1.13$ imply $\nu = 1.41$ and $\chi \approx 1.23$~\cite{PhysRevE.53.414}.

If the decrement of the activated site $\Delta(t)$ is completely added to only the left nearest neighbor, the dynamics becomes anisotropic. The anisotropic version of the model becomes exactly solvable but belongs to a different universality class~\cite{PhysRevLett.75.1550}. In fact, the critical exponents are sensitive to underlying symmetry and dimensionality~\cite{PhysRevE.53.414}. 
For the anisotropic variant of the 1D RH model, $\tau = 4/3$ and $\mathcal{D} = 3/2$ yield $\nu = 2$ and $\chi = 1/2$~\cite{PhysRevLett.75.1550}. 

A correspondence exists between the RH model and the dry friction model~\cite{PhysRevE.74.066110}. In the dry friction model, one can consider two interfaces that move against each other. Let the distance between these two interfaces be $(h_m - h_i)$, where $h_m$ represents the maximum value. The value of $h_m$ fluctuates from the maximum to a value close to the critical height $h_c$. If $h_m$ is high, only the maximum height remains in the contact between the interface, leading to a small frictional force. If $h_m$ is close to $h_c$, many sites come into contact with the interface, so the frictional force is large. The critical height takes a value of nearly $h_c \approx 0.114$~\cite{PhysRevE.74.066110}. The frictional force between two interfaces satisfies a power-law probability distribution $P(F) = F^{-\mu} $, with $\mu = (\mathcal{D}+1/\nu)/{\rm D}$. In the 1D RH model, the exponent is $\mu \approx 3$.

Motivated by recent studies of the $1/f$ noise in the BS model~\cite{PhysRevE.108.044109}, this paper aims to uncover the temporal correlations in the local height or local force noise in the RH model and its variants. Our extensive numerical studies reveal the local force power spectra follow the $1/f^{\alpha}$ noise with a non-trivial value for the spectral exponent. As expected, the spectral exponent changes for the anisotropic variant of the model. The power-law scaling feature is valid for a frequency regime $f\gg f_0$, where the cutoff frequency scales with system size as $f_0 \sim L^{-\lambda}$. We argue that the cutoff frequency exponent is not a new exponent: $\lambda = \mathcal{D}$. In the mean-filed limit, the spectral exponent tends to $\alpha \to 2$ and $\lambda \to 1$. We also examine power spectra for the random walk signal (time evolution of the extremal site) and a local activity signal. Three distinct frequency regimes emerge, and the power spectrum remains system-size-dependent in the entire frequency regime. With finite-size scaling, the critical exponents and scaling function are determined. 

The organization of the paper is as follows: section~\ref{sec_model}
begins with the definition of the RH model. Section~\ref{sec_result} shows
numerical results for the power spectra of the following signals: local force, position of the extremal site, and local activity. The finite-size scaling method reveals the critical exponent and the data collapse. The paper concludes with a summary and discussion in section~\ref{sec_summary}.

\begin{figure}[t]
  \centering
   \scalebox{1}{\includegraphics{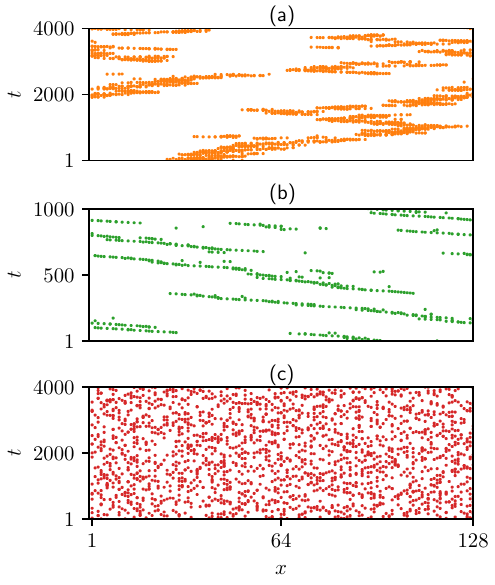}}
    \caption{The space-time evolution of the largest force site for different versions of the model: (a) the RH, (b) the aRH, and (c) the rRH. In the aRH model, the space inversion symmetry $x\to -x$ does not hold.}
  \label{fig_1_pdf}
\end{figure}

\begin{figure}[t]
  \centering
  \scalebox{1}{\includegraphics{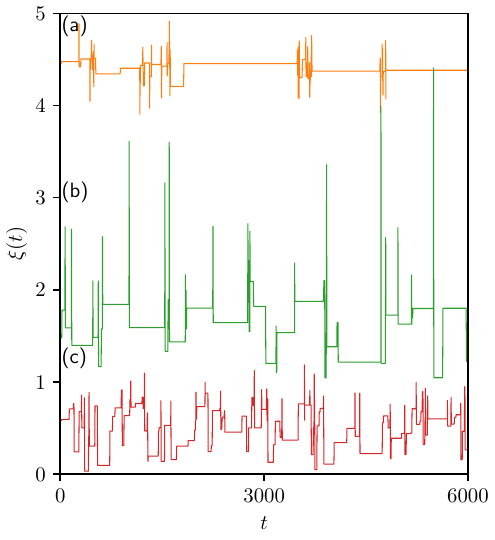}}       
  \caption{The typical temporal noise in the force of a site for (a) the RH, (b) the aRH, and (c) the rRH versions of the model. The local force reveals that the period of stasis may be longer for the RH model than that of the aRH model.}
  \label{fig_2_pdf}
\end{figure}

\section{Model}{\label{sec_model}}
The extremal SOC model that we study basically describes low-temperature creep (popularly known as the Robin Hood system)~\cite{ZAITSEV1992411}. Consider $L$ sites on a circle, where each site has a state variable $\xi$, representing local force. Initially, we assign a random value to each $\xi$ from a uniform distribution $\rho(\xi)$ in a unit interval. The dynamics include the following steps:
\begin{enumerate} 
\item Pick the site with maximum force $\xi_i$. 

\item Reduce a part of the force randomly $\xi_i \to \xi_{i}^{'}$ and transfer that amount $\Delta = \xi-\xi'$ to the nearest neighbors in equal amounts $\xi_{i\pm 1} \to \xi_{i\pm 1}+\Delta/2$.  

\item Goto step 1 and repeat the process ad infinitum.
\end{enumerate}

At each time, the maximum force site triggers an update of the process, implying extremal dynamics.
In the critical state, the extremal site in the space-time plane evolves into a fractal structure (cf figure~\ref{fig_1_pdf}). The position of the extremal site $x(t)$ executes a random walk with a jump size satisfying a power-law distribution. At each update time, we can call the extremal site active. The local activity $A(t)$ takes a value of 1 at time $t$ if the site becomes active and 0 otherwise. 
One of the striking features is that the total force $\eta = \sum_{i=1}^{L}\xi_i$ or the average force $\bar{\xi} = \eta/L$ remains constant during the entire dynamics. 

In this model, the local interaction involves two sites, the nearest left and right neighbors. One can term this an isotropic version of the RH model. Although the model is not solvable, an anisotropic version (aRH) becomes tractable~\cite{PhysRevLett.75.1550}. In the aRH model, the transfer of excess force from the extremal site happens to only the left nearest neighbor. If the addition of force occurs at randomly selected two sites, we term this random neighbor version (rRH). 

Our interest is to examine the temporal correlation in the local force fluctuations $\xi(t)$ (cf figure~\ref{fig_2_pdf}). Using Monte Carlo simulations, we get the signals and compute the power spectral density, employing the standard fast Fourier transform method. In all numerical results, we use a signal of length $N = 2^{18}$ or $2^{20}$ after removing transients up to $10^6$. We also perform ensemble averages over $M = 10^4$ different realizations of the signal and vary the system size from $L = 2^4$ to $2^7$.

\begin{figure}[t]
  \centering
     \scalebox{1.0}{\includegraphics{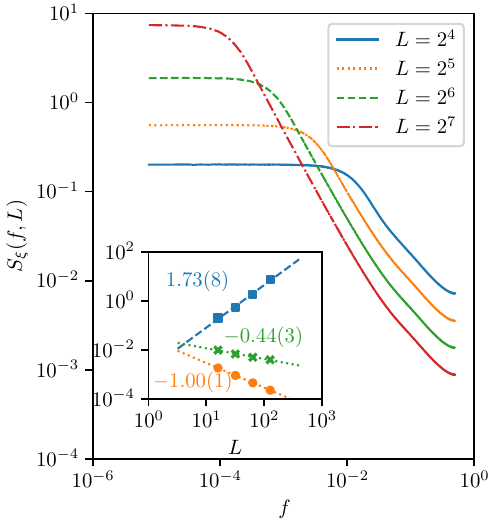}}
  \caption{(Main panel) In the RH model, the power spectra $S_{\xi}(f, L)$ of the local force noise $\xi(t)$  for different system sizes $L$.  (Inset) The system size scaling of various quantities: ({\color{cyan}$\blacksquare$}) the power  below the cutoff frequency $S_{\xi}(L, f\ll f_0)$, ({\color{orange}$\CIRCLE$}) the power  at a fixed frequency above the cutoff frequency $S_{\xi}(L, f=0.1)$, and ({\color{green}$\times$}) the total power $P(L)$, along with the best-fits. The floating numbers [cf table~\ref{tab1}] are the estimated slopes of the straight lines on the double-logarithmic plot.}
  \label{fig_ps_le}
\end{figure}

\begin{figure}
        \centering
 	\scalebox{1.0}{\includegraphics{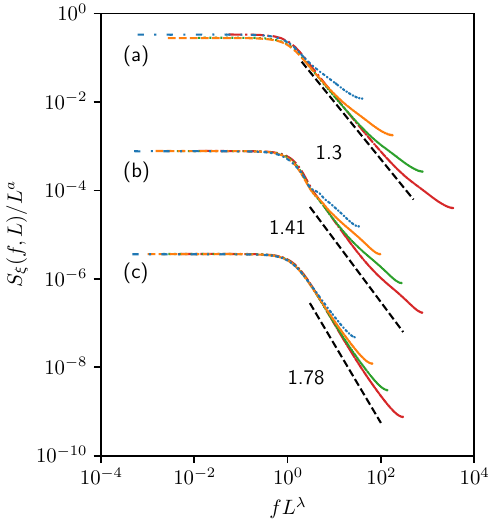}} 
	 \caption{For the local force signals $\xi(t)$, the data collapse $S_{\xi}(f, L)/L^{a}$ with $fL^{\lambda}$ for  (a) the RH, (b) the aRH, and (c) the rRH versions of the model. To compare, we also add straight lines along with the estimated slope values.}		
	\label{fig_Hu_le}
\end{figure}

\begin{table}[t]
	\centering
	\begin{tabular}{|c|ccc|cc|}
		\hline 		
		   Model  & $a$&  $b$ & $a-\lambda$  &$\lambda$  & $\alpha=(a+b)/\lambda$      \\
		\hline
 		RH &  1.73(8) & 1.00(1) & -0.44(3) & 2.2(1) & 1.3(1) \\
 
		aRH & 1.37(1) & 0.80(2) & -0.17(1) & 1.54(2) & 1.41(4)  \\

		rRH & 1.05(1)  & 0.92(2) & -0.06(1) & 1.11(2) & 1.78(6) \\
		\hline
		\end{tabular}
		\caption{The power spectra critical exponents for the local force signal $\xi(t)$.}
	\label{tab1}
\end{table}

\begin{figure}[t]
  \centering
    \scalebox{1}{\includegraphics{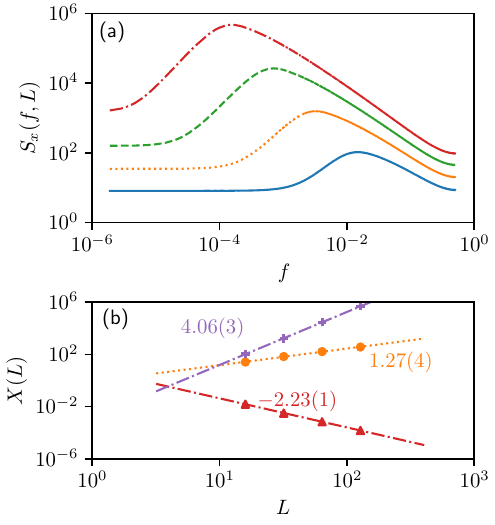}}    
  \caption{(a) The power spectra $S_{x}(f, L)$ for the time series $x(t)$ in the RH model. (b) The system size scaling for ({\color{violet}$+$})  the peak power $S_{x}(L, f=f_p)$, ({\color{red}$\blacktriangle$}) the frequency $f_p$ with peak power, and  ({\color{orange}$\CIRCLE$}) the power  at a fixed frequency above the cutoff frequency $S_{x}(L, f=0.1)$ (cf table~\ref{tab2}).} 
   \label{fig_ps_xt}
\end{figure}

\begin{figure}
\centering
\scalebox{1}{\includegraphics{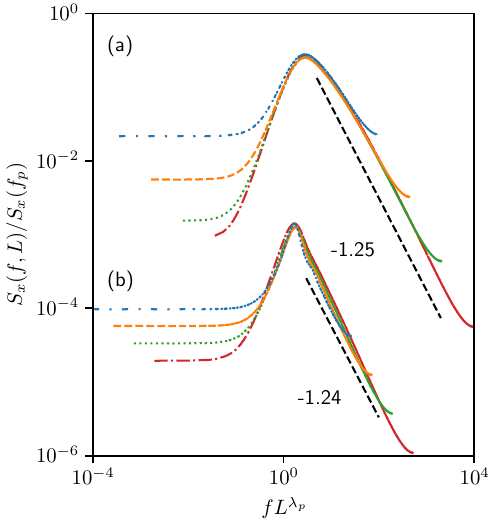}}
   \caption{The scaling function is $S_x(f, L)/S_x(f_p)$, with $f_p \sim L^{-\lambda_p}$ and $S_x(f_p) \sim  L^{a_p}$. The data collapse for the RH (a) and the aRH (b) models.}   
   \label{fig_Hu_xi}
\end{figure}

\begin{table}[t]
	\centering
	\begin{tabular}{|c|cc|cc|}
		\hline 
		 Model   &  $a_p$ & $b$  & $\lambda_p$ & $\alpha=(a_p+b)/\lambda_p$       \\
		\hline
		RH &  4.06(3) & -1.27(4) & 2.23(1) & 1.25(4) \\

		aRH & 3.16(4) & -1.38(1) & 1.43(2) & 1.24(3)  \\
		
		\hline
	\end{tabular}
	\caption{Same as in Table~\ref{tab1}, but for the random walk signal $x(t)$.}
	\label{tab2}
\end{table}

\begin{figure}[t]
  \centering
  \scalebox{1}{\includegraphics{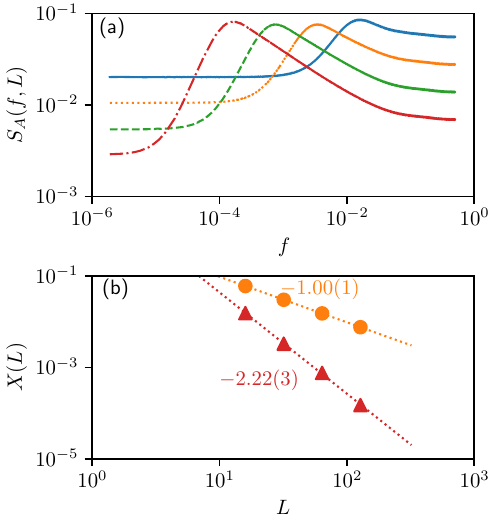}}      
  \caption{(a) In the RH model, the power spectra $S_A(f, L)$ for the local activity signal $A(t)$. (b) The system size scaling for ({\color{red}$\blacktriangle$}) the peak power frequency $f_p$  and  ({\color{orange}$\CIRCLE$}) the power $S_{A}(L, f=0.1)$ at a fixed frequency above the cutoff frequency (cf table~\ref{tab3}).}
  \label{fig_ps_at}
\end{figure}

\begin{figure}
\centering
 \scalebox{1}{\includegraphics{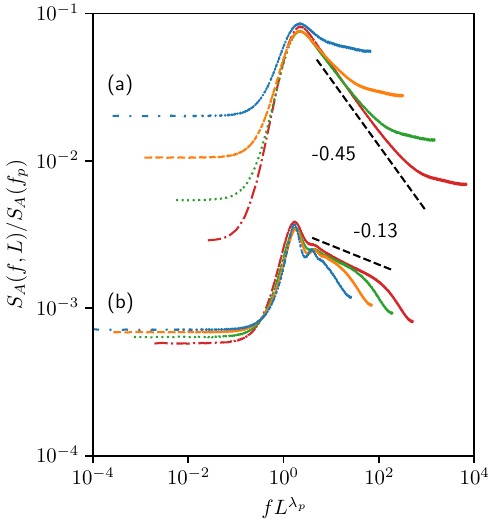}}  
\caption{The data collapse for $A(t)$. The case (a) belongs to the RH model and (b) corresponds to aRH model.}    
 \label{fig_Hu_at}
\end{figure}

\begin{table}[t]
	\centering
	\begin{tabular}{|c|cc|cc|}
		\hline 
		Model  &  $a_p$ & $b$  & $\lambda_p$  &$\alpha=(a_p+b)/\lambda_p$      \\
		\hline
		RH & 0 & 1.00(1) & 2.22(3) & 0.45(1) \\
		
		aRH & -0.81(4) & 0.99(1) & 1.43(2) & 0.13(4)  \\		
		\hline
	\end{tabular}
	\caption{Same as in Table~\ref{tab1}, but for the local activity $A(t)$.}
	\label{tab3}
\end{table}

\section{Results}{\label{sec_result}}
\subsection{The local force noise}
For the RH model, figure~\ref{fig_ps_le} shows the power spectra $S_{\xi}(f, L)$ of the local force noise $\xi(t)$. The spectrum remains independent of frequency below a cutoff frequency of $f_0\sim L^{-\lambda}$ and varies as the $1/f^{\alpha}$ form above $f_0$. On increasing the system size, the power at a fixed frequency increases $\sim L^a$ for $f\ll f_0$ while $\sim L^{-b}$ above the cutoff. One can write an expression for the spectrum as~\cite{PhysRevE.104.064132, Yadav_2022, Kumar_2022, PhysRevE.108.044109}
\begin{equation}
S_{\xi}(f,L) \sim \begin{cases} L^a, ~~~~~~~~~~~~~~~~ f\ll L^{-\lambda},    \\1/\left(f^{\alpha}{ L^{b}}\right), ~~~~~~L^{-\lambda}\ll f \ll 1/2. \end{cases}\nonumber
\end{equation} 
In terms of reduced frequency $u \sim fL^{\lambda}$, the spectrum behaves as
\begin{equation}
S_{\xi}(f,L) \sim L^aH(u),
\label{eq_p2}
\end{equation}
where the scaling function (cf figure~\ref{fig_Hu_le}) varies as $H(u) \sim 1/u^{\alpha}$ for $u\gg1$ and constant  for $u\ll 1$.

Also, the total power scales as
$P(L) \sim L^{a-\lambda}.$
The critical exponents satisfy scaling relations~\cite{PhysRevE.108.044109}
\begin{equation}
 \alpha = (a+b)/\lambda. \nonumber
\end{equation}
It is easy to appreciate the scaling relations in the following way: Since the scaling function $H(u) \sim L^{-a}S_{\xi}(f, L)$ is independent of the system size $L$ in the non-trivial frequency regime, we get $H(u) \sim 1/(f^{\alpha}L^{a+b}) \sim L^{\alpha\lambda}/(L^{\lambda}f)^{\alpha}L^{a+b} \sim 1/u^{\alpha}$, giving $\alpha \lambda = a+b$. Similarly, the total power varies as $P(L) \sim \int df S_{\xi}(f,L) \sim L^a\int df H(u) \sim L^{a-\lambda} $.
As seen from equation~(\ref{eq_p2}), the two exponents $a$ and $\lambda$ determine the scaling function. The exponents are easy to determine from the scaling of the power in low-frequency components and the total power as a function of system size.

\subsection{The random walk and the local activity}
In the RH model, we show power spectra for the random walk and the local activity signals in figure~\ref{fig_ps_xt} and figure~\ref{fig_ps_at}, respectively. The power spectrum does not depend on the frequency below a cutoff frequency $f_0$ and varies in a power-law manner $1/f^{\alpha}$ above a peak frequency $f_p$. Interestingly, the spectrum also shows an intermediate frequency regime $f_0 \ll f\ll f_p$, where the power increases in a power-law manner. The frequency $f_p \sim L^{-\lambda_p}$ corresponds to the peak power $S(f_p) \sim L^{a_p}$. Here, we get the data collapse function (figures~\ref{fig_Hu_xi} and~\ref{fig_Hu_at}) as
\begin{equation}
S(f, L)/S(f_p) = L^{-a_p}S(f, L),\nonumber
\end{equation} 
where the reduced frequency is $fL^{\lambda_p}$. The collapse curve remains independent of $L$ only if $f\gg f_0$, while it decays in a power-law manner $ \sim L^{-(a_p-a)}$ in the frequency regime $f\ll f_0$. In the frequency regime $f\gg f_0$, the scaling function remains independent of $L$, resulting in
\begin{equation}
 \alpha = (a_p+b)/\lambda_p.  \nonumber
\end{equation}

\subsection{Random neighbor version}
For the random neighbor version, the maximum force site and the local activity remain uncorrelated in time. The local force exhibits $\sim 1/f^\alpha$ with $\alpha = 1.8(1)$ very close 2. This behavior seems consistent with mean-field theory.

\section{Summary and Discussion}{\label{sec_summary}}
In summary, we have studied the Robin Hood model in one dimension with different interaction rules. While the interacting sites include the nearest one left and one right neighbor in the original model, an anisotropic case has one left neighbor, and a mean-field version has two random neighbors. Since the Robin Hood system demonstrates self-organized criticality, it is natural to expect the emergence of long-range space-time correlations. In the model, the evolution of the largest force site (random walk) describes a fractal structure in the space-time plane, although the total force remains conserved. 
Specifically, we examined the temporal correlations in the local force noise, the random walk signal, and the local activity. 
By applying the finite-size scaling, we get the data collapse for the power spectra, which reveals the scaling function and the critical exponents. The local force noise follows the $1/f^{\alpha}$ form with the spectral exponent $\alpha \approx 1.3$, and the cutoff frequency varies as $f_0\sim L^{-\lambda}$ with $\lambda \approx 2.2$. In the anisotropic variant, the critical exponents take different values: $\alpha \approx 1.4$ and $\lambda \approx 1.5$.
In the random neighbor version, the local force exhibits mean-field behavior, with $\alpha \approx 1.8$ close to 2 and $\lambda \approx 1.1$ close to 1.

For the random walk and the local activity signals, the power spectrum exhibits three frequency regimes. In the frequency regime below the cutoff frequency $f \ll f_0$, the power spectrum shows constant behavior in the frequency, which depends on the system size. For the intermediate frequency regime $f_0 \ll f\ll f_p$, the power increases in a power law manner, where $f_p$ is the peak frequency corresponding to the maximum power. The power spectrum varies in a decaying power-law manner $1/f^{\alpha}$ above a peak frequency $f\gg f_p$. The power spectrum also shows scaling with the system size in the entire frequency regime. Although the spectral exponent is the same $\alpha \approx 1.2$ for the random walk signals both in the Robin Hood model and its anisotropic variant, the cutoff frequency exponent takes different values of 2.2 and 1.4, respectively. 
For the local activity, the spectral exponent is $\alpha \approx 0.45$ (0.13) for the isotropic (anisotropic) version. Numerically, the two exponents are nearly the same $\lambda \approx \lambda_p$.

Notice that below the cutoff frequency, the noise becomes uncorrelated. The inverse of the cutoff frequency is the average time by which all sites at least once have been updated or lost the retained temporal memory. The system size scaling of these suggests the cutoff frequency exponent is equal to the avalanche dimension $\lambda = \mathcal{D}$. Since the anisotropic version of the model remains exactly solvable with $\mathcal{D} = 3/2$, our numerical estimate of $\lambda$ agrees well~\cite{PhysRevLett.75.1550}. 

Finally, we contrast two extremal SOC models. The Robin Hood model follows the conservation law, where the total force is constant. In the Bak-Sneppen model~\cite{PhysRevE.108.044109}, the sum of state (fitness) variables can fluctuate, implying non-conservative dynamics. 
In both models, the power spectral density for the random walk and the local activity signal exhibits qualitatively different behavior. Three (instead of two) distinct frequency regimes emerge in the Robin Hood model. Particularly in the moderate frequency regime (which is absent in the Bak-Sneppen model), the power increases with frequency and goes up to a peak value at a frequency, which we termed the peak frequency.

\section*{ACKNOWLEDGMENTS}
AS acknowledges Banaras Hindu University for financial support through [Grant No. R/Dev./Sch/UGC Non-Net Fello./2022-23/53315]. ACY recognizes a seed grant under the IoE (Seed Grant-II/2022-23/48729). RC is grateful for the Junior Research Fellowship from UGC, India.

\bibliography{s1sources}
\bibliographystyle{iopart-num}
\end{document}